\documentclass[letterpaper]{sfchem}
\usepackage{graphicx}

\begin{document}

%\title{ A clumpy model for Photon Dominated Regions: Effects of metallicity}
\title{Metallicity effects in Photon Dominated Regions: Clumpy clouds}

\author{S. Jeyakumar and J. Stutzki} 
  \institute{
I. Physikalisches Institut,
Universit\"at zu K\"oln,
Z\"ulpicher Strasse 77,
50937 K\"oln, Germany.
} 
%Short author list here: Surnames only please (no initials)
\authorrunning{Jeyakumar and Stutzki}
%Short title here:
\titlerunning{PDRs and metallicity effects}

\maketitle 

\begin{abstract}

Several galaxies, such as dwarfs and Irregulars as well as outer galactic clouds 
have low metallicity. At low metallicities a reduction in the amount of dust and heavy
elements plays a significant role on the chemistry as well as the heating and cooling of
the gas in the molecular regions, called as Photon Dominated Regions (PDRs).
We present here the effects of reduced metallicity in PDRs and 
study the important PDR cooling lines ([CII], [CI] and CO).
Moreover many observational evidences suggest that 
molecular clouds are clumpy.
%The clump spectrum analysis of the molecular clouds show a power law
%spectrum of mass. 
We model the molecular emission from galaxies incorporating 
a mass spectrum of clumps.  We also compare our results with the semi-analytical results 
obtained by Bolatto et al. (1999).


\end{abstract}

\section{Introduction}

Photon Dominated Regions are predominantly molecular and atomic regions where
the physical and chemical processes are dominated by UV radiation
(cf. Hollenbach \& Tielens, 1997, 1999). 
The molecular clouds in the vicinity of the newly formed stars is heated by the
FUV photons in the energy range from about 6 to  13.6eV. These clouds cool through
the atomic and molecular spectral lines, such as [CII]158$\mu m$,  [OI]63,146$\mu m$, 
[CI]609,370$\mu m$ and the milli-metric and sub-mm CO rotational lines. 
Plane parallel and spherical models of PDRs have been constructed to understand the
physical and chemical characteristics of these regions
(e.g., Kaufman et al., 1999; K\"oster et al., 1994; Le Bourlot et al., 1993; Sternberg \& Dalgarno 1995; 
St\"orzer et al., 1996; Tielens \& Hollenbach 1985). However there are other important factors which affect the 
UV absorption and scattering as well as the basic heating and cooling processes in PDRs.

Several galaxies such as Dwarf galaxies, Irregular galaxies, the Large and Small Magellenic
Clouds have low metallicity (cf. Wilson 1995). 
A radial gradient in metallicity of molecular clouds is found in 
the Milky Way and several other nearby galaxies (e.g., Arimoto et al., 1996).  
These low metallicity systems have much higher [CII]/CO and [CI]/CO line ratios   
as compared to the galactic value (e.g. Bolatto et al., 2000; Madden 2000). 
This suggests that the effects of metallicity should be considered while interpreting 
the molecular and atomic spectral line observations of these sources.
Since the important surface coolant of the PDRs, the [CII]158$\mu$m emission, is strongly 
affected by the metallicity factor, we study the effects of metallicity in PDRs. 
In addition, low metallicity PDR models would also help us to understand 
the star forming regions in Dwarf galaxies which resemble the primordial galaxies.

Additionally, observations of edge-on PDR have suggested that the molecular 
clouds are clumpy, and the UV radiation can penetrate deep inside the clouds 
(cf. Stutzki et al., 1988; Boiss\'e 1990). 
This suggests that a simple single component model may not be sufficient to explain the 
observed features. We also include a mass spectrum of clumps to understand the
cooling lines of PDRs from low metallicity galaxies. 

\section{Metallicity effects}

Low metallicity systems have a lower, dust to gas ratio and heavy elemental abundances 
as compared to the galactic ISM. 
This reduction in the amount of dust grains affects, 
the absorption of UV radiation, heating of gas by photo electric emission from dust, 
formation of H$_2$ on the dust grains and the cooling of the gas through atomic and 
molecular lines. In addition, the chemistry is also affected by the reduction
in dust and heavy elements (van Dishoeck \& Black 1988; Lequeux et al., 1994).
We use the self-consistent spherical PDR model of St\"orzer et al. (1996) to study
the metallicity effects. We scale the dust dependent parameters and the abundance of
heavy elements with the metallicity factor, Z, in our PDR calculations. 

%\subsection{Surface temperature}

\begin{figure}[h]
\resizebox{\hsize}{!}{
\includegraphics[clip=]{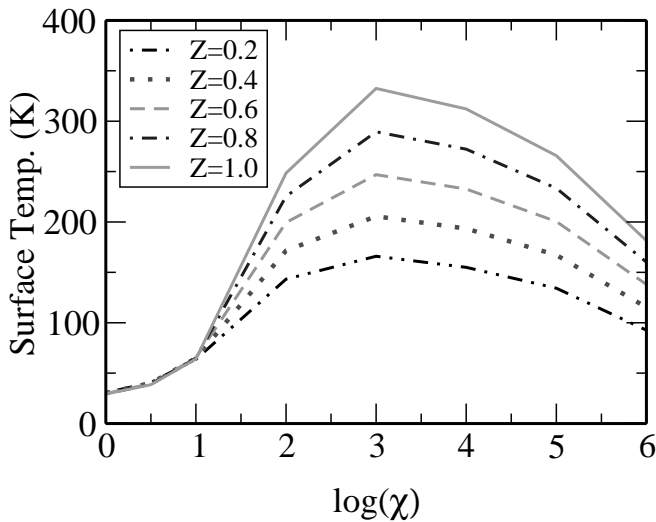}}
\caption{The temperature at the surface of the PDR clump of mass 
$M=1~M_{\odot}$ and density $n = 10^3 cm^{-3}$ is plotted against the UV 
field for different metallicities. \label{pdrtemp} }
\end{figure}

The variation of the temperature at the surface of the PDR clumps with the incident UV field 
is plotted in figure~\ref{pdrtemp}, for a clump of mass, M=1M$_\odot$ and density, 
n=10$^3cm^{-3}$.  The UV field, $\chi$, is expressed in units of mean UV field of Draine (1978).
It is seen from the figure that at high UV fields the temperature is 
proportional to metallicity, whereas at low UV fields there is no significant 
change in the surface temperature.

This correlation can be understood analytically
by balancing the dominant heating and cooling processes. 
In the case of PDRs exposed to high UV fields grain photo-electric emission (PE) dominates 
the heating. 
The photo electric heating rate given by Bakes \& Tielens (1994) is
\begin{eqnarray}
\Gamma_{PE}  = 10^{-24} \chi n Z {3 \times e^{-2} \over { 1 + 2\times10^{-4}{\chi T^{1/2} \over n_e} } }
\label{PEheat}
\end{eqnarray}
where Z is the metallicity factor.

\begin{figure}[h]
\resizebox{\hsize}{!}
{\includegraphics[clip=]{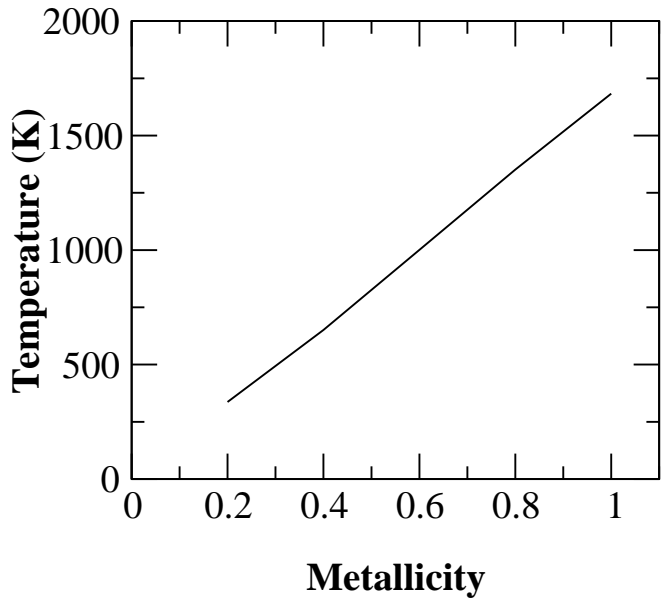}}
\caption{The equilibrium temperature for the balance between grain photo-electric heating
rate and [OI] cooling rate is plotted against the metallicity, for a density, n=10$^{5.5}cm^{-3}$
and UV field, $\chi=10^6$. \label{equitemp} }
\end{figure}

The cooling is dominated by fine structure [OI] emission, gas-grain collisions and 
fine-structure [CII] emission.  The cooling rate can be written as
$L_{ij} = n_i A_{ij} h\nu_{ij} \beta{(\tau_{ij})}$ where $\beta$ is the escape probability,
$A_{ij}$ is the transition probability and $n_i$ number of atoms at level $i$ and $h\nu_{ij}$
is the corresponding frequency of radiation.

Following Hollenbach \& McKee (1979), for the population of
the first level of the OI atom, the cooling rate of [OI]63$\mu$m can be written as,
%\begin{eqnarray*}
%{n_1 \over n_{O} } = { 1 \over { 1 + g_0/g_1 e^{E10/kT} + g_2/g_1 e^{-E21/kT} } }
%\end{eqnarray*}
%Thus the cooling rate of [OI]63$\mu$,

\begin{eqnarray}
L_{OI} = { {8.45\times10^{-22} \beta{(\tau)} n_H Z } \over
          { 1  + 5e^{228/T} + 1/3e^{-98/T} } }.
\label{OIcool}
\end{eqnarray}

Although eqns.~\ref{PEheat} and \ref{OIcool} show a similar dependence with metallicity,
the grain heating rate has an additional dependence on metallicity through the 
charge state of the grains. 
The charge state of the grains is expressed as ${\chi T^{1/2} \over n_e}$ which depends
on the availability of electrons for recombination. 
The main source of electrons at high UV fields is the ionisation of CI. 
Since at the surface almost all of the CI is ionised, the electron density 
$n_e \sim n_C \propto n_{C,Z=1}Z$. With this assumption 
%and the carbon abundance at 
%solar metallicity of $1.4\times10^{-4}$, 
the heating rate, $\Gamma_{PE}/Z$ decreases with decreasing $Z$
whereas the cooling rate, $L_OI/Z$ remains constant. 
The equilibrium temperature obtained by solving eqns.~\ref{PEheat} and \ref{OIcool} 
decreases as the metallicity decreases (cf. figure~\ref{equitemp}) as seen in our PDR 
calculations.

%\subsubsection{Width of CII and CI layer}
% The widths of the CII and CI layers are plotted against the metallicity
%for a clump of mass M=1$M_\odot$ and density $n_0=10^4 cm^{-3}$ (figure on the left).
%This figure shows that the width of the CII $\propto 1 /Z$.
%This is due to the reduced absorption of UV by the dust as the metallicity decreases.
%However the width of the CI region saturates to a constant value as the metallicity decreases
%implying a chemical origin of CI rather than photo-dissociation of CO.
%
%\begin{figure}[h]
%\resizebox{\hsize}{!}{\includegraphics{width_CI_CII.ps}}
%\caption{Width}
%\end{figure}
%

\section{Clumpy PDR model} 
\label{clumpymodel}

In our clumpy model, the molecular cloud is modelled as being composed of many spherical 
clumps of mass spectrum of the form, 
%\begin{eqnarray*}
$\frac{dN}{dM} = A M^{-\alpha}.$
%\label{massspectrum}
%\end{eqnarray*}
We use $\alpha = 1.8$ (e.g. Kramer et al., 1998).
We assume that the turbulent velocity dispersion of the cloud is larger than the intrinsic line
width of each clump. Thus the clumps do not interact radiatively and the total 
intensity of a spectral line from the cloud can be written as,
\begin{eqnarray*}
I_{T} = \int_{M_{min}}^{M_{max}} I(M) \left(\frac{dN}{dM}\right)_{cloud} ~ f_{b,c} ~ f_{M,b} dM
\end{eqnarray*}
The beam filling factor of each clump is $f_{M,b}(=\Omega_M/\Omega_b$) where $\Omega_M$ is the
solid angle of the clump of mass $M$ and $\Omega_b$ is the beam solid angle. The fraction of the clumps within the beam is given by
$f_{b,c}= \Omega_b/\Omega_c$.

By scaling the mass with $M_{max}$ ($x = M/M_{max}$)
\begin{center}
$I_{T} = A M_{max}^{1-\alpha}~ f_{b,c} ~f_{M_{max},b} ~\int_{M_{min}/M_{max}}^{1} I_{x} x^{-\alpha} f_{M,M_{max}} dx$
\end{center}

\begin{eqnarray*}
A \approx  M_{tot} (2 -\alpha) M_{max}^{-(2-\alpha)}
\label{constA}
\end{eqnarray*}
where $f_{b}$ is the beam filling factor, $x_0= x(M_{min})$ and
$f_x = R^2(x)/R^2_{cloud}$.
Although the total intensity depends on the scaling constant and the beam filling factors, 
the line ratio between any two spectral lines depends only on 
$M_{min}/M_{max}$ and $\alpha$.

\section{Clumpy clouds at low metallicity} 

%In low metallicity clumps, the UV radiation is less absorbed due to a reduced amount of 
%dust grains, thus producing a large layer of CII by ionising CI and a small CO core deep 
%inside the clump. 
The observations of [CII] and [CI] emission from low metal galaxies have been modelled
by Bolatto et al. (1999), assuming that the size of the CII region scales inversely with 
metallicity. It is also assumed that the size of the CI region remains between two 
limiting scenarios of, an inverse dependence and no variation with metallicity. 
Our spherical PDR model calculations based on the model of  St\"orzer et al. (1996) show 
that the size of the CII layer is indeed inversely proportional to the metallicity factor Z.
However the size of the CI region shows very weak dependence at low Z and a roughly 
inverse dependence at high Z for a typical spherical clump of density, $10^4 cm^{-3}$. 

\begin{figure}[h]
\resizebox{\hsize}{!}{\includegraphics{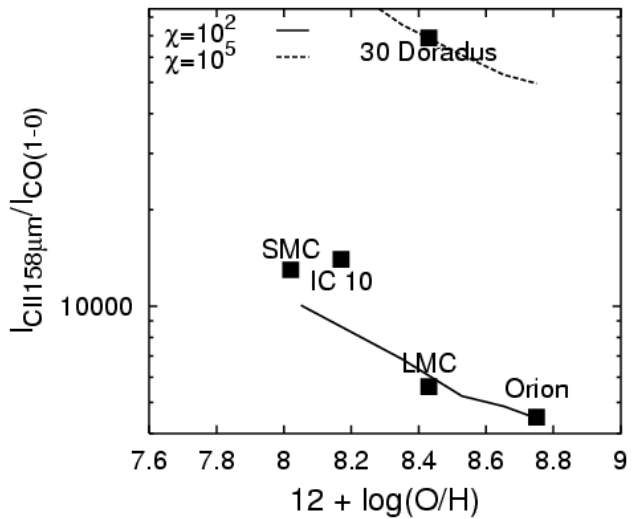}}
\caption{
The intensity ratio $I_{[CII]158\mu m}$/$I_{CO(1-0)}$ 
is plotted against the metallicity. The maximum and minimum masses of the clump 
spectrum are 10$^3$ and 10$^{-2}$ M$_\odot$.  The density at the surface of the 
clumps are 10$^4 cm^{-3}$. The observed ratios of nearby galaxies are plotted 
as filled squares. The model result plotted as solid line compares well 
with that of the trend shown in figure~5  of Bolatto et al. (1999). 
\label{clumpyplot}
}
\end{figure}

The intensity ratio $I_{[CII]158\mu m}$/$I_{CO(1-0)}$, observed in many nearby 
low metal galaxies, show a power law dependence with metallicity. This dependence 
has been predicted using a semi-analytical clumpy model by  Bolatto et al. (1999).
We use our clumpy model explained in section~\ref{clumpymodel} to study the 
metallicity dependence of this line ratio.
%The individual clump intensities for different metallicities
%are derived from the PDR calculations based on the spherical model of St\"orzer et al. (1996).
The observed ratios of nearby galaxies can be well represented
by a clumpy model of density $10^4~cm^{-3}$, exposed to a UV field of $\chi=10^2$
as shown in figure~\ref{clumpyplot}. The higher observed ratio for the 30 Doradus region 
can be explained by a similar clumpy model, but exposed to an UV field of  $\chi=10^5$. 
These results compare well with the results shown by the semi-analytical model of 
Bolatto et al. (1999).  This trend suggests that at low metallicities  CII is a tracer of 
molecular hydrogen rather than CO.  However the observed variation of  
[CI]/CO line ratio  with metallicity is steeper than the model prediction. 
%This is due to large [CI] line intensities predicted by the model and requires further study of the PDR models to explain the observations.
This is most likely due to large [CI] line intensities predicted by the PDR models
and requires further investigation.
%and requires further study of the PDR models to explain the observations.

\begin{acknowledgements}
The research presented here is supported by the Deutsche Forschungsgemeinschaft
(DFG) via Grant SFB\,494. 
\end{acknowledgements}

\end{document}